\documentstyle[epsf]{kapproc}

\kluwerbib
\let\lcitebracket(
\let\rcitebracket)
\startauthorindex

\begin{document}

\articletitle{Binary Population Synthesis: Methods, Normalization, and
Surprises}
 
\author{Vassiliki Kalogera}
\affil{Harvard-Smithsonian Center for Astrophysics, Cambridge, MA
02138, USA}
\email{vkalogera@cfa.harvard.edu}

\author{Krzysztof Belczynski}
\affil{Harvard-Smithsonian Center for Astrophysics, Cambridge, MA
02138, USA; Nicolaus Copernicus Astronomical Center, 00-716 Warszawa,
Poland}
\email{kbelczynski@cfa.harvard.edu}

\begin{abstract} 

 In this paper we present a brief overview of population synthesis
methods with a discussion of their main advantages and disadvantages. In
the second part, we present some recent results from synthesis models of
close binary compact objects with emphasis on the predicted rates, their
uncertainties, and the model input parameters the rates are most
sensitive to. We also report on a new evolutionary path leading to the
formation of close double neutron stars (NS), with the unique
characteristic that none of the two NS ever had the chance to be recycled
by accretion. Their formation rates turn out to be comparable to or maybe
even higher than those of recycled NS--NS binaries (like the ones
observed), but their detection probability as binary pulsars is much
smaller because of their short lifetimes. We discuss the implications of
such a population for gravitational--wave detection of NS--NS inspiral
events, and possibly for gamma--ray bursts and their host galaxies.

\end{abstract}

\section{INTRODUCTION}

Binary systems with compact objects (neutron stars or black holes) have
provided us with some of the most surprising stellar configurations
discovered in galaxies: from high-- and low--mass X-ray binaries,
persistent or transient and mini--quasars to binary millisecond radio
pulsars and double neutron stars. For decades a lot of effort has been
devoted to understanding the evolutionary history of these systems,
identifying the dominant formation processes and factors that determine
their properties. This multi--faceted research effort has followed two
main directions. One includes studies of specific observed systems with
the goal to understand their characteristics and origin, and provide us
with clues about the general population properties. Although it is
possible to tailor an evolutionary model to reproduce the properties of
an isolated system, this exercise provides little perspective on whether
the putative initial conditions and subsequent tailoring are plausible. A
different and more general approach is to model the evolution of an
entire ensemble of primordial binaries under a common set of assumptions.
Such {\em population synthesis} models (e.g., Dewey \& Cordes 1987;
Politano 1988; Lipunov \& Postnov 1988; de Kool 1992; Kolb 1993; and many
more) can be very useful in analyzing the statistical properties of the
population under study and allows comparisons to observed samples of
objects.

In the first part of the paper, we discuss in some detail two different
methods used in binary population synthesis calculations and the reasons
for choosing one or the other (\S\,2). In the second part, we focus on
some recent results we have obtained from studies of the formation of
X-ray binaries and double compact objects, without attempting at any
level to review this area of research (\S\,3, 4). In our discussion of
the results we focus on issues related to the absolute normalization of
models and predicted formation rates, ways of constraining them and their
sensitivity to model input parameters. In \S\,4, we report on a new class
of double neutron star systems with implications for gravitational--wave
detection and gamma-ray bursts.

\section{Population Synthesis Methods} 

The basic goal in population synthesis calculations is to follow the
evolution of an ensemble of primordial binaries through all possible
evolutionary phases until the formation of the systems of interest.
Typical examples of such phases include: wind mass loss, stable or
unstable and conservative or non--conservative mass and angular momentum
transfer phases, formation of compact objects with associated mass loss
and supernova kicks, circularization, angular momentum loss through
gravitational radiation and/or magnetic braking. The details of some of
these phases is not well understood, but in general the choice of
prescriptions and assumptions about their treatment in the models is
guided by detailed hydrodynamic or stellar evolution studies. Most often
for the modeling of stellar evolution and the physical processes
involved, analytic approximations adequate for statistical estimates are
employed.

In what follows we discuss in some more detail two very different methods
that have been used so far in population synthesis calculations.

\subsection{Evolution of Distribution Function in Phase Space}

A class of semi-analytical methods have been employed for population
synthesis calculations in various flavors over the years (e.g., Kolb 1993;
Politano 1996; Kalogera \& Webbink 1998; Kalogera 1998). These methods are
based on the idea of evolving a distribution function through phase space
using Jacobian transformations. In the specific case of binaries, the
``phase space'' is the space of binary parameters, i.e., masses, orbital
separations, and eccentricities. A distribution function describing the
population of binaries at the beginning of a given evolutionary phase,
$F(M_1,M_2,A,e)$, can be transformed into another function
$F'(M'_1,M'_2,A',e')$ describing the binaries at the end of this phase.
The transformation can be performed using Jacobians (involves partial
derivatives of final binary parameters with respect to initial parameters)
and provided that the functional relationships between final and initial
quantities are known:
 \begin{equation}
F\left(M'_1,M'_2,A',e'\right)~=~F'\left(M_1,M_2,A,e\right)\,\left[\cal{J}\,
\left(\frac{M'_{{\rm 1}},M'_{{\rm
2}},A',{\rm e}'}{M_{{\rm 1}},M_{{\rm 2}},A,{\rm e}}\right)\right]^{-1}.
 \end{equation}
 In the vast majority of cases in binary evolution, these relationships
are not just known but they are such that the partial derivatives can be
calculated analytically! It is often the case that the evolution of a
distribution function can be calculated analytically through a whole
sequence of evolutionary phases (e.g., wind mass loss, common envelope,
stable mass transfer, supernova explosions). Some of the mathematical
details of such cases have been derived and discussed, for example, by
Kalogera (1996) and Kalogera \& Webbink (1998). In its general form this
computational method is {\em semi-analytical} because at times it is
required that integrations are performed over physical parameters that are
not of interest. Typically such integrations can be performed only
numerically.

These methods of phase--space evolution have great advantages in providing
us with high--accuracy results, high resolution in distributions of binary
parameters at low computational costs. The numerical implementation is
done at the final stage of evolution, after the binary population has been
evolved through a given evolutionary sequence analytically. The grid over
which the distribution function is calculated is set up on the actual set
of parameters and evolutionary stage of interest and the density of this
grid can be chosen to be high enough so that numerical noise is not an
issue. This is to be contrasted with the concept of setting up a ``grid''
on the initial parameters of the population and anticipating that this
initial grid will be dense enough to describe well the final properties of
the population.  In addition to high numerical accuracy and low
computational cost, the semi-analytical population synthesis method offers
physical insight to how the final properties of the population depend on
the properties of their progenitors and on the physical parameters that
enter the description of the various evolutionary phases. This invaluable
physical insight originates from the analytical derivation of the
distribution functions in each evolutionary stage and it allows us to
identify the dominant physical quantities that dictate the final results.
Often it also allows us to derive analytical, strict limits of physical
properties (e.g., Kalogera 1996; Kalogera \& Lorimer 2000) or functional
dependencies between parameters.

There is however one major disadvantage to the phase-space evolution
method: it can be used only if the exact evolutionary sequence is known  
{\em a priori}. In other words, the calculation of the evolution (in
practice of the Jacobian transformations) has to be ``tailored'' to a
specific path and, for populations of systems that are formed through a   
multitude of evolutionary paths, the analysis has to be repeated for each 
one of them.  Another consequence is that population studies performed    
solely with this method cannot lead to the discovery of new unexplored  
channels of evolution.

\subsection{Monte Carlo Techniques}

Monte Carlo methods have a straightforward application in population
synthesis calculations and has been used most often in the study of binary
systems (e.g., Dewey \& Cordes 1987; Lipunov \& Postnov 1988;
Portegies--Zwart \& Verbunt 1996; Han, Podsiadlowski, \& Eggleton 1995;
and many others). The basic design relies on the idea of generating a
large number of unevolved binary and single stars with physical properties
generated according to pre-determined distributions. The evolution of each
``Monte Carlo object'' is followed through a wide variety of evolutionary
stages until the formation of systems of interest (i.e., binary compact
objects). The evolution modeling relies on assumptions and prescriptions
for relevant physical processes, in a way very similar to phase--space
evolution calculations.

Probably the most important disadvantage of Monte Carlo techniques in
population synthesis studies is that of statistical accuracy of the
results. Even with current computational resources, statistical accuracy
can still be a problem for studies of rare populations with relatively low
formation rates, such as binary compact objects and low--mass X--ray
binaries. Reducing statistical noise in predicted rates of double neutron
stars, for example, to less than even 10\% requires a total number of
primordial binaries in excess of $\sim 10^6$ and poses very serious
computational--cost requirements (typically 100 hours on fast
workstation). Consequently, achieving a satisfactory statistical accuracy
for multi-dimensional distribution of rare populations over physical
parameters and not just their rate needs careful examination of the models
and top--level computational resources. With respect to gaining physical
insight to the properties of the population of interest and what
determines them, Monte Carlo techniques are less suitable than
phase--space evolution methods. For example, a strict limit on physical
characteristics cannot be {\em rigorously} identified in Monte Carlo
simulations. Typically only statistical statements can be made and there
is always the question whether a certain area of the parameter space is
physically excluded or just underpopulated.

On the other hand, among the most important advantages of Monte Carlo
techniques used in population studies are: (i) they are rather simple in
their conceptual design and implementation, (ii) numerical calculations
can be parallelized in a straightforward manner, (iii) they allow the
study of a large number of evolutionary sequences and stellar populations
simultaneously, and (iv) they can lead to the discovery of {\em new},
previously unappreciated or ignored formation channels for objects of
interest (see \S\,4).

\section{Population Synthesis of Double Compact Objects: Predicted Rates}

In what follows we discuss preliminary results from a Monte Carlo
population study of double compact objects, neutron stars (NS) or black
holes (BH), forming through a multitude of evolutionary channels
(Belczynski, Kalogera, \& Bulik 2001). In particular we focus on the
absolute normalization of the models and the predicted formation rates of
close NS--NS, BH--NS, and BH--BH binaries. Interest in these formation
rates arises from the importance of such close systems for
gravitational--wave astronomy and possibly for gamma--ray bursts. The late
inspiral of close binary compact objects are primary sources for the
ground--based laser interferometers currently under construction (such as
LIGO, VIRGO, GEO600), and predictions of detection rates depend
sensitively on the formation rates of these objects. On the other hand,
the question of the origin of gamma--ray bursts (GRB) and their possible
connection to compact object mergers can to be addressed to some extent 
through a comparison of GRB and binary compact object formation rates.

To address the above questions, though, knowledge of theoretical formation
rates are essential. Population synthesis models, however, often lack a
good calibration and predicted merger rates available in the literature
tend to cover a large range (many orders of magnitude; for a recent
review see Kalogera 2001). Therefore, it is important to use a number of
ways to better constrain the absolute normalization of synthesis models.

\subsection{Observational Constraints} 

For the calibration of the population synthesis models for double compact
object formation presented here, we have used a number of independent
constraints on stellar populations derived from observations: (i) rate of
Type II supernovae thought to be core collapse events of hydrogen--rich
massive stars, (ii) rate of Type Ib/c thought to be core collapse events
of hydrogen--poor stars, (iii)  star formation history for our Galaxy,
(iv) empirical merger rates of NS--NS binaries obtained based on the
observed NS--NS sample. Additional constraints can also be obtained from
(i) observed frequency of helium stars in isolation and in binaries, (v)
frequency and lifetime of compact object binaries with helium stars, and
(vi) formation rates of X--ray binaries.

This list of constraints includes populations and events that are directly
related to the formation of double compact objects but also some others
that are not, but their formation can be followed (as side products) in
population synthesis calculations. We use the recent determination of
supernova rates of Cappellaro, Evans, \& Turatto (1999). The derived rates
for Type II and Ib/c supernovae for a galaxy like the Milky Way lead to a
rate ratio (II/Ibc) of $6 \pm 4$. Estimates of the Milky Way star
formation rate lie in the range $1-3$\,M$_\odot$\,yr$^{-1}$ (Blitz 1997;  
Lacey \& Fall 1985). Empirical estimates of the NS--NS merger rate when
taking all uncertainties into account see to lie in the range
$10^{-6}-10^{-4}$\,yr$^{-1}$ (Kalogera et al.\ 2000).

\subsection{Model Assumptions} 

Here, we give a brief description of our population synthesis code. More
details about the treatment of various evolutionary processes will be
presented in Belczynski, Kalogera, \& Bulik (2001).

To describe the evolution of single or non--interacting binary stars
(hydrogen-- and helium--rich) from the zero age main sequence (ZAMS) to
carbon--oxygen (CO) core formation, we employ the analytical formulae of
Hurley, Pols, \& Tout (2000), whose results are in good agreement with
earlier stellar models (e.g., Schaller et al.\ 1992). To calculate masses
of compact objects formed at core--collapse events, we have adopted a
prescription based on the relation between CO core masses and final FeNi
core masses (Woosley 1986). Our progenitor--remnant mass relation is in
agreement with the results of Fryer \& Kalogera (2001) based on
hydrodynamic calculations of core collapse of massive stars.

Concerning the evolution of interacting binaries, we model the changes of
mass and orbital parameters (separation and eccentricity) taking into
account mass and angular momentum transfer between the stars or loss from
the system during Roche--lobe overflow, tidal circularization,
rejuvenation of stars due to mass accretion, wind mass loss from massive
and/or evolved stars, dynamically unstable mass transfer episodes leading
to common--envelope (CE) evolution and spiral--in of the stars.  We extend
the usual treatment of CE evolution based on energy considerations
(Webbink 1984) to include cases where both stars have reached the giant
branch and have convective envelopes (hydrogen or low--mass helium stars).
As suggested by Brown (1995) for hydrogen--rich stars, we expect the two  
cores to spiral--in until a merger occurs or the combined stellar
envelopes are ejected. We also account for the {\em possibility} that
compact objects accrete mass during CE phases (following Brown 1995). At
NS formation, we model the effects of asymmetric supernovae (SN)  on
binaries, i.e., mass loss and natal kicks, for both circular and eccentric
orbits (e.g., Kalogera 1996; Portegies--Zwart \& Verbunt 1996). We assume 
that kicks are isotropic with a given magnitude distribution.

In the synthesis calculations, we evolve a population of primordial
binaries and single stars through a large number of evolutionary stages,
until {\em coalescing} double compact objects are formed (merger times
$<10$\,Gyr). The total number of binaries (typically a few million) in
each simulation is determined by the requirement that the statistical
(Poisson) fractional errors ($\propto 1/\sqrt N$) of the final populations
are lower than 10\%.

In our standard model, the properties of primordial binaries follow
certain assumed distributions: for primary masses ($5-100$\,M$_\odot$),
$\propto M_1^{-2.7}dM_1$; for mass ratios ($0<q<1$), $\propto dq$; for
orbital separations (from a minimum, so both ZAMS stars fit within
their Roche lobes, up to $10^5$\,R$_\odot$), $\propto dA/A$; for      
eccentricities, $\propto 2e$. Each of the models is also characterized by
a set of assumptions, which, for our standard model, are:
 \begin{itemize}
 \item{{\em Kick velocities.} We use a weighted sum of two Maxwellian
distributions with $\sigma=175$\,km\,s$^{-1}$ (80\%) and
$\sigma=700$\,km\,s$^{-1}$ (20\%) (Cordes \& Chernoff 1997); Black--hole
kicks are weighted based on the amount of fallback mass during their
formation.} 
 \item{{\em Maximum NS mass.} We adopt a conservative value of $M_{\rm
max}=3$\,M$_\odot$ (e.g., Kalogera \& Baym 1996). It affects the relative
fractions of NS and BH and the outcome of NS hyper--critical
accretion in CE phases.}
 \item{ {\em Common envelope efficiency.} We assume $\alpha_{\rm
CE}\times\lambda = 1.0$, where $\alpha$ is the efficiency with which
orbital energy is used to unbind the stellar envelope, and $\lambda$ is a
measure of the central concentration of the giant.}
 \item{ {\em Non--conservative mass transfer.} In cases of dynamically
stable
mass transfer between non--degenerate stars, we allow for mass and angular
momentum loss from the binary (see Podsiadlowski, Joss, \& Hsu 1992),
assuming that half of the mass lost from the donor is also lost from the
system ($1-f_{\rm a}=0.5$) with specific angular momentum equal to
$\beta2\pi$A$^2$/P ($\beta=1$).}
 \item{{\em Star formation history.} We assume that star formation has
been continuous in the disk of our Galaxy for the last 10\,Gyr (e.g.,
Gilmore 2001).}
 \end{itemize}

An extensive parameter study is essential in assessing the robustness of
population synthesis results. Apart from our standard case, we examine the
results for 25 additional models, where we vary all of the above
parameters within reasonable ranges. The complete set of models and the
assumptions that are different from our standard choices are: 
 \begin{itemize}
 \item{A: standard model} 
 \item{B1--7: zero kicks, single Maxwellian with
$\sigma=50,100,200,300,$\\ $400$\,km\,s$^{-1}$,
``Paczynski'' kicks with $\sigma=600$\,km\,s$^{-1}$} 
 \item{C: no hyper--critical accretion onto NS and BH in CEs}
 \item{D: maximum NS mass: $M_{\rm max}=2, 1.5$\,M$_\odot$} 
 \item{E1--6: $\alpha_{\rm CE}\times\lambda = 0.1, 0.25, 0.5, 0.75, 2, 3$} 
 \item{F1--4: mass fraction accreted: f$_{\rm a}=0.1, 0.25, 0.75, 1$} 
 \item{G1--2: specific angular mom.\ $\beta=0.5, 2$} 
 \item{H: primary mass: $\propto M_1^{-2.35}$} 
 \item{I1--2: mass ratio: $\propto {q}^{-2.7}$, $\propto {q}^{3.0}$} 

 \end{itemize}

\subsection{Most Important Model Parameters and their Effects} 

In assessing the sensitivity of our results to the model input parameters
varied as described in \S\,3.2, we examine the behavior of (i) the
calculated ratio of Type II to Type Ib/c rates, and (ii) the formation
rates of three types of coalescing double compact objects: NS--NS,
BH--NS, and BH--BH. 

We find that the SN rate ratio is by far most sensitive to the assumed
binary fraction for the stellar population. In agreement with earlier
results (de Donder \& Vanbeveren 1998), the ratio increases with
decreasing binary fraction, primarily because the majority of Type Ib/c
supernovae occur in binary systems, where loss of the hydrogen envelope is
facilitated. This behavior tentatively points to a low binary fraction for
our Galaxy (about 25\%), or conversely to the galaxies used to determine
the SN rates having a binary fraction lower than that in the Milky Way. We
should note however, that the predicted SN rate ratios are consistent with
the observational estimates within the one-sigma error bar.

Variations of the mass ratio distribution appear to be the most important
factor for the predicted coalescence rates. A distribution that
strongly favors high mass ratios leads to an increase of the NS--NS rate
by a factor of about 30 relative to the rate in the case that low mass
ratios are strongly favored. The BH--NS and BH--BH rates are less
sensitive and show variations by factors of about 3 and 6, respectively. 

The common--envelope efficiency is well known to be an important model
parameter in population synthesis. Rates of NS binaries decrease by about
an order of magnitude, for a similar decrease of the product $\alpha_{\rm
CE}\lambda$, which leads to an increased rate of merger prior to the
formation of the binaries of interest. However, the effect on the
coalescence rate of BH--BH binaries is non-monotonic. The reason is that
these binaries tend to form with wide orbits. Decreasing the CE
efficiency initially causes the rate to {\em increase}, as more tight
binaries are allowed to form. With further decrease of the efficiency
though, the early merger rate dominates and the formation rate of BH--BH
binaries drops. 

Last, the effect of kicks (imparted to NS but also to BH with smaller
magnitudes) affects the double compact object formation in many different
ways: supernova survival, formation of tight binaries, mergers. The
formation typically peak for $\sigma = 100-200$\,km\,s$^{-1}$ and
decrease with increasing average kick magnitudes. The rate of close
BH--NS systems is found to vary the most in our parameter study by a
factor of about 3.

\section{New Class of Close NS--NS: Without NS Recycling}

Empirical rate estimates of NS--NS mergers, are derived under an important
underlying assumption that all NS--NS binaries have contained a rather
long--lived radio pulsar at some point in their lifetime (see Kalogera et
al.\ 2000).  However, in our very recent Monte Carlo population synthesis
calculations (Belczynski \& Kalogera 2000), we discovered a {\em new}
evolutionary path leading to NS--NS formation that does {not} involve
recycling of any of the two neutron stars. The implication of these
findings is that there may be a significant population of old NS--NS
binaries that could not be detected as binary pulsars. We use the
calculated formation rate of non--recycled NS--NS relative to that of
recycled NS--NS systems to derive upward revisions for the rates estimates
based on the current observed sample.

\subsection{Model Calculations and Results}

We use our population synthesis models to investigate all possible
formation channels of NS--NS binaries realized in the simulations. We find
that a significant fraction of {\em coalescing} NS--NS systems are formed
through a new, previously not identified evolutionary path.

\begin{figure}
\epsfbox{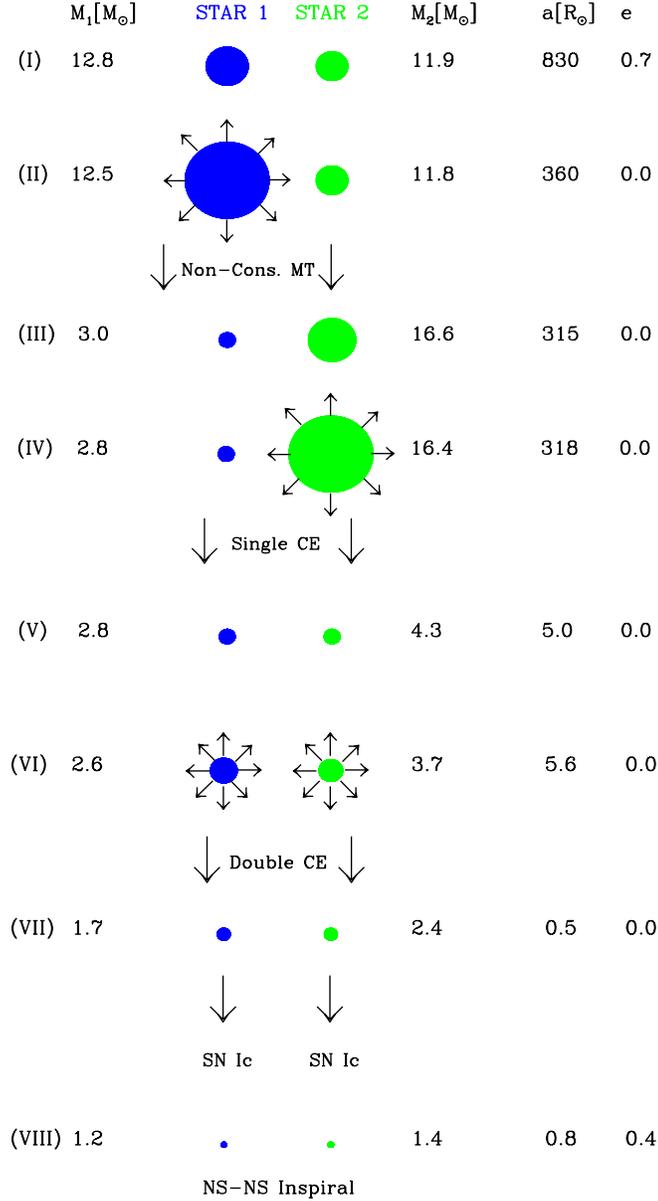}
\caption{Stages of the new non--recycled NS--NS formation path:
    (I) Zero Age Main Sequence,
    (II) star 1 fills its Roche lobe and non--conservative mass transfer
begins,
    (III) at the end of stage II the helium core of Star 1 is
exposed,
    (IV) star 2 fills its Roche lobe on the giant branch leading to
dynamically unstable mass transfer and CE evolution, 
    (V) at the end of stage IV the helium core of Star 2 is
exposed,
    (VI) both helium stars fill their Roche lobes on the giant
branch, leading to a double CE phase,
    (VII) at the end of stage VI, the binary consists of two
          bare CO cores in a tight orbit
    (VIII) after two subsequent supernovae and 20\,Myr since ZAMS
a close NS--NS binary forms with a merger time of about 5\,Myr. (From
Belczynski \& Kalogera 2000.)
}
\end{figure}

In Figure 1 we describe in detail the formation of a typical NS--NS binary
through this new channel. The evolution begins with two phases of
Roche--lobe overflow. The first, from the primary to the secondary,
involves non--conservative but dynamically stable mass transfer (stage II)
and ends when the hydrogen envelope is consumed. The second, from the
initial secondary to the helium core of the initial primary, involves
dynamically unstable mass transfer, i.e., CE evolution (stage IV). The
post--CE binary consists of two bare helium stars of relatively low
masses. As they evolve through core and shell helium burning, the two
stars acquire ``giant--like'' structures, with developed CO cores and
convective envelopes (e.g., Habets 1987). Their radial expansion
eventually brings them into contact and the system evolves through a
double CE phase (stage VI; similar to Brown (1995) for hydrogen--rich
stars). During this double CE phase, the combined helium envelopes are
ejected at the expense of orbital energy. The tight, post--CE system
consists of two CO cores, which eventually end their lives as Type Ic
supernovae. The survival probability after the two supernovae is quite
high, given the tight orbit before the explosions. The end product in this
example is a close NS--NS with a merger time of $\simeq 5$\,Myr (typical
merger times are found in the range $10^4-10^8$\,yr).

The unique qualitative characteristic of this NS--NS formation path is
that both NS have avoided recycling. The NS progenitors have lost both
their hydrogen and helium envelopes prior to the two supernovae, so no
accretion from winds or Roche--lobe overflow is possible after NS
formation. Consequently, these systems are detectable as radio pulsars
only for a time ($\sim 10^6$\,yr) much shorter than recycled NS--NS pulsar
lifetimes ($\sim 10^8-10^{10}$\,yr in the observed sample). Such short
lifetimes are of course consistent with the number of NS--NS binaries
detected so far and the absence of any {\em non--recycled} pulsars among
them.

Given the uncertain absolute normalization of population synthesis models,
we focus primarily on the formation rate of non--recycled NS--NS binaries
{\em relative} to that of recycled pulsars, formed through other,
qualitatively different evolutionary paths. Based on this comparison, for
each of our models, we derive a correction factor for empirical estimates
of the Galactic NS--NS coalescence rate. Since these estimates are derived
based on the observed sample, they can account only for NS--NS systems
with recycled pulsars, and they must be increased to include any
non--recycled systems formed. For the majority of the examined models, we
find this correction factor to lie in the range $1.5-3$ although for some
models it can be as high as 10 or even higher (for more details see
Belczynski \& Kalogera 2000). 

As already mentioned, the realization of the newly identified path through
a double helium CE phase depends on the final stages of a helium star
evolution. It has long been known that low mass helium stars, after core
helium exhaustion, expand significantly and develop a ``giant-like''
structure with a clearly defined core and a convective envelope 
(Habets 1987; Avila-Reese 1993; Woosley, Langer, \& Weaver
1995; Hurley et al.\ 2000). We further examined in detail models of
evolved helium stars (Woosley 1997, private communication) and found that
helium stars below 4.0\,M$_\odot$ have deep convective envelopes and that
slightly more massive helium stars ($\sim$ 4--4.5\,M$_\odot$) still form
convective envelopes although shallower. Evolved stars with convective
envelopes, overfilling Roche lobes in binary systems, transfer mass on a
dynamical time scale, and as a consequence CE evolution ensues. The
development of CE phases was proposed first in the context of cataclysmic
variable formation (Paczynski 1976) and is now supported by detailed
hydrodynamic calculations in a variety of binary configurations (e.g.
Rasio \& Livio 1996; Taam \& Sandquist 2000 and references therein). At
present no hydro calculations exist for the case of two evolved stars.  
Based on our basic understanding of CE development, it seems reasonable to
expect that, if two stars with convective envelopes are involved in a mass
transfer episode, a double core spiral-in can occur leading to double CE
ejection (Brown 1995).  Based on these earlier calculations, we adopt a
maximum helium--star mass for CE evolution (double or single) of
4.5\,M$_\odot$. The formation rates of both types of NS--NS binaries are
somewhat sensitive to this value, because they depend on whether helium
stars evolve through CE phases (single and double, for recycled and
non-recycled systems, respectively).  Reducing the maximum mass to
4\,M$_\odot$ actually increases the rate correction factor, although by
less than 20\%, as it reduces the recycled NS--NS rate by a factor larger
than the non--recycled rate.

We note that the identification of the formation path for non--recycled
NS--NS binaries stems entirely from accounting for the evolution of helium
stars and for the possibility of double CE phases, both of which have
typically been ignored in previous calculations (with the exception of   
Fryer et al.\ 1999, although formation paths for non--recycled NS--NS were
not discussed).

\subsection{DISCUSSION}

We have identified a new possible evolutionary path leading to the
formation of close NS--NS binaries, with the unique characteristic that   
both NS have avoided recycling by accretion. The realization of this path
is related to the evolution of helium--rich stars, and particularly to the
radial expansion and development of convective envelopes (on the giant
branch) of low--mass ($< 4.5$\,M$_\odot$) helium stars.
We find that a significant fraction of coalescing
NS--NS form through this new channel, for a very wide range of model
parameters. In some cases, the non--recycled NS--NS systems strongly 
dominate the total close NS--NS population.

Since both NS are non--recycled, their pulsar lifetimes are too short (by
$\sim 10^3$), and hence their detection probability is negligible relative
to recycled NS--NS. However, intermediate progenitors of these systems may
provide evidence in support of the evolutionary sequence. Examples are
close binaries with two low--mass helium stars or a helium star with an
O,B companion. Although there are selection effects against their 
detection too (e.g., short lifetimes, high--mass helium stars are
brighter, broad spectral lines due to winds, high luminosity contrast
between binary members, etc.), it has been estimated that so far only
about 1\% of binary helium stars have been detected (Vrancken et al.\ 
1991). Therefore, it is reasonable to expect that the observed sample will
increase in future years as observational techniques improve. It is worth
noting that theoretical calculations indicate that explosions of low--mass
helium stars formed in binaries reproduce the light curves of type Ib
supernovae (Shigeyama et al.\ 1990).

The results of our population modeling show that, if one accounts for the
formation of non--recycled NS--NS binaries, the total number of coalescing
NS-NS systems could be higher by factors of at least 50\%, and up to 10 or
even higher. Such an increase has important implications for prospects of
gravitational wave detection by ground--based interferometers. Using the
results of Kalogera et al.\ (2000) on the empirical NS--NS coalescence
rate, we find that their {\em most optimistic} prediction for the LIGO I
detection rate could be raised to at least 1 event per 2--3 years, and
their {\em most pessimistic} LIGO II detection rate could be raised to
3--4 events per year or even higher.

Our results also have important implications for gamma--ray bursts
(GRBs), if they are associated with NS--NS coalescence. We find that
the typical merger times of the non--recycled NS--NS are considerably
shorter than those of recycled binaries, whereas their center--of--mass
velocities are higher. The balance of these two competing effects could
alter the current consensus for the location of GRB progenitors
relative to their host galaxies (e.g., Belczynski, Bulik, \& Rudak
2000; Bloom, Kulkarni, \& Djorgovski 2000; Fryer et al.\ 1999).

\begin{acknowledgments}
 Support is acknowledged by the Smithsonian Institution through a CfA
Predoctoral Fellowship to KB and a Clay Fellowship to VK, and by a Polish
Nat.\ Res.\ Comm.\ (KBN) grant 2P03D02219 to KB.

\end{acknowledgments}

%{%\normallatexbib

%\bibliographystyle{apalike}
%\chapbblname{chapbib}
%\chapbibliography{logic}

%}


\begin{chapthebibliography}{1}

\bibitem{}
Avila-Rees, V.\ A.\ 1993, Rev.\ Mex.\ Astron.\ Astrofis., 25,
79

\bibitem{}
Belczynski, K., Bulik, T., \& Rudak, B.\ 2000, A\&A, 
submitted [astro-ph/0011177]

\bibitem{}
Belczynski, K., Kalogera, V.\ 2000, ApJ Letters, submitted
[astro-ph/0012172]

\bibitem{}
Belczynski, K., Kalogera, V.\ \& Bulik T., 2001
in preparation

\bibitem{}
Blitz, L.\ 1997, in ``CO: Twenty--Five Years of Millimeter-Wave
Spectroscopy'', eds.\ W.B.\ Latter, et al.\ (Kluwer Academic Publishers),
11


\bibitem{}
Bloom, J.S., Kulkarni, S.R., \& Djorgovski, S.G.\
2000, ApJ, submitted [astro-ph/0010176]

\bibitem{}
Brown, G.E.\ 1995, ApJ, 440, 270

\bibitem{}
Cappellaro, E., Evans, R., \& Turatto, M.\ 1999, ApJ, 351, 459

\bibitem{}
Cordes, J., \& Chernoff, D.F.\ 1997, ApJ, 482, 971

\bibitem{}
de Donder, E., \& Vanbeveren, D.\ 1998, A\&A, 333, 557

\bibitem{}
de Kool, M. 1992, A\&A, 261, 188

\bibitem{}
Dewey, R. J., \& Cordes, J. M. 1987, ApJ, 321, 780

\bibitem{}
Fryer, C.L., \& Kalogera, V.\ 2001, ApJ, in press [astro-ph/9911312]

\bibitem{}
Fryer, C.\ L., Woosley, S.\ E., \& Hartmann, D.\ H.\
1999, ApJ, 526, 152

\bibitem{}
Gilmore, G.\ 2001, to appear in ``Galaxy Disks and Disk  
Galaxies'', eds.\ J.G.\ Funes \& E.M.\ Corsini (San Francisco: ASP)

\bibitem{}
Habets, G.\ M.\ H.\ J.\ 1987, A\&A Supplement Series, 69, 183

\bibitem{}
Han, Z., Podsiadlowski, P., \& Eggleton, P.P.\ 1995, MNRAS, 272, 800
 
\bibitem{}
Hurley, J.R., Pols, O.R., \& Tout, C.A.\ 2000, MNRAS, 315, 543

\bibitem{}
Kalogera, V.\ 1996, ApJ, 471, 352

\bibitem{}
Kalogera, V.\ 1998, ApJ, 493, 368

\bibitem{}
 Kalogera, V.\ 2001, to appear in {\em Workshop on Astrophysical Sources
for Ground-Based Gravitational Wave Detectors}, ed.\ J.\ Centrella

\bibitem{}
Kalogera, V., \& Baym, G.A.\ 1996, ApJ, 470, L61   

\bibitem{}
Kalogera, V., \& Lorimer, D.R.\ 2000, ApJ, 530, 890

\bibitem{}
Kalogera, V., Narayan, R., Spergel, D.N., \&
Taylor, J.H.\ 2000, ApJ, submitted [astro-ph/0012038]

\bibitem{}
Kalogera, V., \& Webbink, R.F.\ 1998, ApJ, 493, 351

\bibitem{}
Kolb, U. 1993, A\&A, 271, 149

\bibitem{}
Lacey, C.G., \& Fall, S.M.\ 1985, ApJ, 290, 154

\bibitem{}
Lipunov, V. M., \& Postnov, K. A. 1988, Astrophysics \& Space Science,
145, 1

\bibitem{}
Paczynski, B.\ 1976, in ``The Structure and Evolution of
Close Binary Systems'', eds. P.\ Eggleton, S.\ Mitton, \& J.\ Whelan,
Dordrecht, Reidel, p.\ 75

\bibitem{}
Podsiadlowski, P., Joss, P.C., \& Hsu, J.J.L.\ 1992,
ApJ, 391, 246

\bibitem{}
Politano, M. J. 1996, ApJ, 465, 338

\bibitem{}
Portegies--Zwart, S.\ F., \& Verbunt, F.\ 1996, A\&A,
309, 179

\bibitem{}
Portegies--Zwart, S.\ F., \& Yungel'son, L.\ R.\ 1998,
A\&A, 332, 173

\bibitem{}
Rasio, F.\ A., \& Livio, M.\ 1996, ApJ, 471, 366

\bibitem{}
Schaller, G., Schaerer, D., Meynet, G., \&
Maeder, A. 1992, A\&A Supplement Series, 269, 331

\bibitem{}
Shigeyama, T., Nomoto, K., Tsujimoto, T., \&
Hashimoto, M.\ 1990, ApJ, 361, L23

\bibitem{}
Tamm, R.\ E., \& Sandquist, E.\ L.\ 2000, ARA\&A,
38, 113

\bibitem{}
Vrancken, J., DeGreve, J.P., Yungel'son, L., \&
Tutukov, A.\ 1991, A\&A, 249, 411

\bibitem{}
Webbink, R.F.\ 1984, ApJ, 277, 355

\bibitem{}
Woosley, S.E.\ 1986, in ``Nucleosynthesis and Chemical
Evolution'', 16th Saas-Fee Course, eds. B. Hauck et al., Geneva Obs., p.\
1

\bibitem{}
Woosley, S. E., Langer, N., \& Weaver, T. A. 1995, ApJ, 448, 315


\end{chapthebibliography}
\end{document}